\documentstyle[twoside]{article}

\catcode`\@=11
\long\def\@makefntext#1{
\protect\noindent \hbox to 3.2pt {\hskip-.9pt  
$^{{\eightrm\@thefnmark}}$\hfil}#1\hfill}               

\def\@makefnmark{\hbox to 0pt{$^{\@thefnmark}$\hss}}    
        
\def\ps@myheadings{\let\@mkboth\@gobbletwo
\def\@oddhead{\hbox{}
\rightmark\hfil\eightrm\thepage}   
\def\@oddfoot{}\def\@evenhead{\eightrm\thepage\hfil
\leftmark\hbox{}}\def\@evenfoot{}
\def\sectionmark##1{}\def\subsectionmark##1{}}

\oddsidemargin=\evensidemargin
\newcounter{sectionc}\newcounter{subsectionc}\newcounter{subsubsectionc}
\renewcommand{\section}[1] {\vspace{12pt}\addtocounter{sectionc}{1} 
\setcounter{subsectionc}{0}\setcounter{subsubsectionc}{0}\noindent 
        {\tenbf\thesectionc. #1}\par\vspace{5pt}}
\renewcommand{\subsection}[1] {\vspace{12pt}\addtocounter{subsectionc}{1} 
        \setcounter{subsubsectionc}{0}\noindent 
        {\bf\thesectionc.\thesubsectionc. {\kern1pt \bfit #1}}\par\vspace{5pt}}
\renewcommand{\subsubsection}[1] {\vspace{12pt}\addtocounter{subsubsectionc}{1}
        \noindent{\tenrm\thesectionc.\thesubsectionc.\thesubsubsectionc.
        {\kern1pt \tenit #1}}\par\vspace{5pt}}

\newcounter{appendixc}
\newcounter{subappendixc}[appendixc]
\newcounter{subsubappendixc}[subappendixc]
\renewcommand{\thesubappendixc}{\Alph{appendixc}.\arabic{subappendixc}}
\renewcommand{\thesubsubappendixc}
        {\Alph{appendixc}.\arabic{subappendixc}.\arabic{subsubappendixc}}

\renewcommand{\appendix}[1] {\vspace{12pt}
        \refstepcounter{appendixc}
        \setcounter{figure}{0}
        \setcounter{table}{0}
        \setcounter{lemma}{0}
        \setcounter{theorem}{0}
        \setcounter{corollary}{0}
        \setcounter{definition}{0}
        \setcounter{equation}{0}
        \renewcommand{\thefigure}{\Alph{appendixc}.\arabic{figure}}
        \renewcommand{\thetable}{\Alph{appendixc}.\arabic{table}}
        \renewcommand{\theappendixc}{\Alph{appendixc}}
        \renewcommand{\thelemma}{\Alph{appendixc}.\arabic{lemma}}
        \renewcommand{\thetheorem}{\Alph{appendixc}.\arabic{theorem}}
        \renewcommand{\thedefinition}{\Alph{appendixc}.\arabic{definition}}
        \renewcommand{\thecorollary}{\Alph{appendixc}.\arabic{corollary}}
        \renewcommand{\theequation}{\Alph{appendixc}.\arabic{equation}}
        \noindent{\tenbf Appendix \theappendixc #1}\par\vspace{5pt}}
\newcommand{\subappendix}[1] {\vspace{12pt}
        \refstepcounter{subappendixc}
        \noindent{\bf Appendix \thesubappendixc. {\kern1pt \bfit #1}}
        \par\vspace{5pt}}
\newcommand{\subsubappendix}[1] {\vspace{12pt}
        \refstepcounter{subsubappendixc}
        \noindent{\rm Appendix \thesubsubappendixc. {\kern1pt \tenit #1}}
        \par\vspace{5pt}}

\newcommand{\smalllineskip}{\baselineskip=10pt}

\def\eightcirc{
\begin{picture}(0,0)
\put(4.4,1.8){\circle{6.5}}
\end{picture}}
\def\eightcopyright{\eightcirc\kern2.7pt\hbox{\eightrm c}}

\renewenvironment{thebibliography}[1]
        {\frenchspacing
         \ninerm\baselineskip=11pt
         \begin{list}{\arabic{enumi}.}
        {\usecounter{enumi}\setlength{\parsep}{0pt}     
         \setlength{\leftmargin 12.7pt}{\rightmargin 0pt} 
         \setlength{\itemsep}{0pt} \settowidth
        {\labelwidth}{#1.}\sloppy}}{\end{list}}

\newcounter{itemlistc}
\newcounter{romanlistc}
\newcounter{alphlistc}
\newcounter{arabiclistc}

\newcommand{\fcaption}[1]{
        \refstepcounter{figure}
        \setbox\@tempboxa = \hbox{\footnotesize Fig.~\thefigure. #1}
        \ifdim \wd\@tempboxa > 5in
           {\begin{center}
        \parbox{5in}{\footnotesize\smalllineskip Fig.~\thefigure. #1}
            \end{center}}
        \else
             {\begin{center}
             {\footnotesize Fig.~\thefigure. #1}
              \end{center}}
        \fi}

\newcommand{\tcaption}[1]{
        \refstepcounter{table}
        \setbox\@tempboxa = \hbox{\footnotesize Table~\thetable. #1}
        \ifdim \wd\@tempboxa > 5in
           {\begin{center}
        \parbox{5in}{\footnotesize\smalllineskip Table~\thetable. #1}
            \end{center}}
        \else
             {\begin{center}
             {\footnotesize Table~\thetable. #1}
              \end{center}}
        \fi}

\def\@citex[#1]#2{\if@filesw\immediate\write\@auxout
        {\string\citation{#2}}\fi
\def\@citea{}\@cite{\@for\@citeb:=#2\do
        {\@citea\def\@citea{,}\@ifundefined
        {b@\@citeb}{{\bf ?}\@warning
        {Citation `\@citeb' on page \thepage \space undefined}}
        {\csname b@\@citeb\endcsname}}}{#1}}

\newif\if@cghi
\def\cite{\@cghitrue\@ifnextchar [{\@tempswatrue
        \@citex}{\@tempswafalse\@citex[]}}
\def\citelow{\@cghifalse\@ifnextchar [{\@tempswatrue
        \@citex}{\@tempswafalse\@citex[]}}
\def\@cite#1#2{{$\null^{#1}$\if@tempswa\typeout
        {IJCGA warning: optional citation argument 
        ignored: `#2'} \fi}}

\def\pmb#1{\setbox0=\hbox{#1}
        \kern-.025em\copy0\kern-\wd0
        \kern.05em\copy0\kern-\wd0
        \kern-.025em\raise.0433em\box0}

\def\fnt#1#2{\footnotetext{\kern-.3em
        {$^{\mbox{\scriptsize #1}}$}{#2}}}

\font\tenrm=cmr10
\font\tenit=cmti10 
\font\tenbf=cmbx10
\font\bfit=cmbxti10 at 10pt
\font\ninerm=cmr9

\font\eightrm=cmr8

\def\qed{\hbox{${\vcenter{\vbox{                        
   \hrule height 0.4pt\hbox{\vrule width 0.4pt height 6pt
   \kern5pt\vrule width 0.4pt}\hrule height 0.4pt}}}$}}

  \def\newpic#1{}

\def\beq{\begin{equation}}
\def\eeq{\end{equation}}

\begin{document}

\begin{titlepage}

\title{Fusion rules for Quantum Transfer Matrices as a 
Dynamical System on Grassmann Manifolds}

\author{O.Lipan \thanks{James Franck Institute of the University of Chicago,
5640 S.Ellis Avenue, Chicago, IL 60637, USA}
\and P.Wiegmann \thanks{James Franck Institute and
and Enrico Fermi Institute of the University of Chicago, 5640 S.Ellis
Avenue, Chicago, IL 60637, USA and
Landau Institute for Theoretical Physics}
\and A. Zabrodin
\thanks{Joint Institute of Chemical Physics, Kosygina str. 4, 117334,
Moscow, Russia and ITEP, 117259, Moscow, Russia}}

\maketitle

\begin{abstract}

We show that the set of transfer matrices of an arbitrary fusion
type for an integrable quantum model obey these bilinear functional
relations, which are identified with an integrable dynamical system
on a Grassmann manifold (higher Hirota equation).  The bilinear
relations were previously known for a particular class of transfer
matrices corresponding to rectangular Young diagrams.  We extend
this result for general Young diagrams.  A general solution of the
bilinear equations is presented.
\end{abstract}

\vfill

\end{titlepage}

\section{Introduction}

One of the key objects in the theory of quantum
integrable systems is the family of transfer matrices ($T$-matrices).
They  are operators acting in the Hilbert space of the quantum model and
represent a commutative set of integrals of motion. 

A $T$-matrix for a quantum model on a one-dimensional lattice is
constructed out of a Lax operator $L_{i,Y}(u)$, where $i$ is a lattice
site and $u$ is a spectral parameter.  The Lax operator is
characterized by a Young diagram $Y$ and acts in the tensor product
$V_i\otimes V_Y$ of the Hilbert space on the site $i$ and the space of
the representation $Y$ (the auxiliary space). The $T$-matrix is then a
trace of the product of Lax operators along the chain, taken over the
auxillary space: \beq { T}^{(Y)}(u)=\mbox{tr}_YL_{N,Y}(u-v_N)\ldots
L_{2,Y}(u-v_2) L_{1,Y}(u-v_1).
 \label{transfer} \eeq
It follows from the Yang-Baxter equation
that these operators
commute for different Young diagrams and
spectral parameters: $[T^{(Y_1 )}(u_1 ), \, T^{(Y_2 )}(u_2 )]=0$.

The $T$-matrices are linearly independent but obey a set 
of functional relations called {\it fusion rules}.
The fusion procedure allows one to construct Lax operators
 and $T$-matrices for higher Young diagrams out of a set of 
Lax operators for lower ones and, ultimately, out of the one-box diagram.
(For the basic material on the
quantum inverse scattering method and the fusion procedure we
refer the reader to the
papers \cite{FT},\,\cite{fusion}, respectively.)

The fusion rules are especially simple and are closed for {\it
  rectangular} Young diagrams \cite{KP},\,\cite{Kuniba}.  Let
$T_{a,s}(u)$ be the transfer matrix for a rectangular diagram with
height $a$ and length $s$. Then the fusion rule may be written in
bilinear form \beq T_{a,s}(u+1)T_{a,s}(u-1) -T_{a,s+1}(u)T_{a,s-1}(u)
=T_{a+1,s}(u)T_{a-1,s}(u).
\label{3term}
\eeq This relation has been identified \cite{KLWZ} with Hirota's
celebrated bilinear difference equation \cite{Hirota1} for the
$\tau$-function for the hierarchy of classical difference integrable
equations.  Since $T$-matrices commute at different $u,\,a,\,s$, the
same relation holds for their eigenvalues, so one can treat $T$ in eq.
(\ref{3term}) as number-valued functions.  These relations can be
treated as an integrable {\it classical} equation and it proves to be
useful to obtain a complete solution of the {\it quantum} problem
\cite{KLWZ}.

The bilinear relation (\ref{3term}) reveals
 an intimate connection between the fusion 
rules and the geometry of Grassmann manifolds. 
The point is that the Hirota equation can be viewed as
 a particular case of the general Pl\"ucker relations for 
coordinates on a Grassmannian manifold \cite{Sato}.

In this Letter we extend this result and show that the fusion rules
for general Young diagrams (not necessarily rectangular) have this
bilinear form and are equivalent to the general Pl\"ucker relations.
These are equivalent to higher Hirota equations -- the hierarchy of
discrete integrable equations. We restrict ourselves to the series
$A_{k-1}$.

The structure of the paper is as follows.  In Sec.\,2, we outline the
higher Hirota equations. In Sec.\,3, we present the fusion relations
in the bilinear form for an arbitrary Young diagram and identify them
with the higher Hirota equation (\ref{H2}). In Sec.\,4 the general
solution to eq.\,(\ref{HH}) is given for a relevant class of boundary
conditions. Finally, Sec.\,6 contains the proofs based on the
Pl\"ucker relations, reviewed in Sec.\,5.

\section{Higher Hirota equations}

The difference soliton equations form {\it hierarchies}
\cite{Miwa2,OHTI} in complete analogy with differential soliton
equations.  Higher members of the hierarchy (higher Hirota difference
equations) are bilinear relations for a function $\tau (x_1 , x_2 ,
\ldots , x_r )$ of $r$ variables.  They have the form \cite{OHTI} \beq
\left | \begin{array}{llllllll}
    1& z_1 & z_{1}^{2}&&\ldots &&z_{1}^{r-2}& \tau _1 \hat \tau _{1}\\
    &&&&&&&\\
    1& z_2 & z_{2}^{2}&&\ldots &&z_{2}^{r-2}& \tau _2 \hat \tau _{2}\\
    &&&&&&&\\
    \ldots &\ldots &\ldots &&\ldots && \ldots &\ldots \\
    &&&&&&&\\
    1& z_r & z_{r}^{2}&&\ldots &&z_{r}^{r-2}& \tau _r \hat \tau _{r}
\end{array}\right | =0\,,
\label{H1}
\eeq
where the $z_i$ are arbitrary constants and
\begin{eqnarray}
&&\tau _{i}\equiv \tau (x_1 , x_2 ,\ldots ,x_{i-1}, x_{i}+1, x_{i+1},
\ldots , x_r )\,, \nonumber\\
&&\hat \tau _{i}\equiv \tau
(x_1 +1, x_2 +1,\ldots ,x_{i-1}+1, x_{i}, x_{i+1}+1,
\ldots , x_r +1)\,.
\label{H1a}
\end{eqnarray}
In a more compact form they read
\beq
\sum _{l=1}^{r} \alpha _{l}\tau _{l}\hat \tau _{l}=0\,,
\label{H2}
\eeq where the $\alpha _l$ are minors of the matrix (\ref{H1}). They
may be treated as independent constants subject to the condition $\sum
_{l=1}^{r}\alpha _{l}=0.$ For $r=3$ one gets the three-term Hirota
equation (\ref{3term}).

The transformation
\beq
\tau (x_1 ,\ldots , x_r )\rightarrow
\exp \left [ \frac{1}{2r-4}\sum _{l=1}^{r}\log \alpha _{l}
\left (\sum _{j=1, \ne l}^{r}x_{j}\right )^{2}
\right ]
\tau (x_1 ,\ldots , x_r )
\label{H3}
\eeq
sends eq.\,(\ref{H2}) to the canonical form,
\beq
\sum _{l=1}^{r} \tau _{l}\hat \tau _{l}=0\,,
\label{H4}
\eeq
with no extra parameters.

\section{Bilinear Fusion Rules}

Let us define
the Young diagram  by the
coordinates of its corners (see the Figure):
\begin{eqnarray}
&&
Q_i =
(\lambda _i,\mu _{i-1}), \;\;\;\;\;\;\; i=1, \ldots , n+1,
\nonumber \\
&&P_i=(\lambda _i ,\mu _i ), \;\;\;\;\;\;\;\;\;\;\; i=1, \ldots , n.
\label{h1}
\end{eqnarray}
We  imply that $\lambda _{n+1}=\mu _{0}=0$.
The coordinates $\lambda _{i}$ and $\mu _{i}$ are ordered in strictly
decreasing and strictly increasing order respectivelly: 
$\lambda_i>\lambda_{i+1}$, $\mu_i<\mu_{i+1}$.
The set of {\it ladder coordinates} is also useful:
\begin{eqnarray}
&&s_i=\lambda _i -\lambda _{i+1} , \;\;\;\;\;\;\;  i=1,\ldots ,n,
\nonumber \\
&&a_i=\mu _{i} -\mu _{i-1} , \;\;\;\;\;\;\; i=1, \ldots ,n.
\label {h2}
\end{eqnarray}
The Young diagram $Y=Y(\lambda, \mu)$, $\lambda =(\lambda _{1}, \ldots
, \lambda _{n})$, $\mu =(\mu _{1}, \ldots , \mu _{n})$ has $a_1$ rows
of length $\lambda _{1}$, $a_2$ rows of length $\lambda _{2}$ and so
on. Hereafter we denote $T^{(Y)}(u)=T_{\lambda}^{\mu}(u)$.


\vspace{3mm}

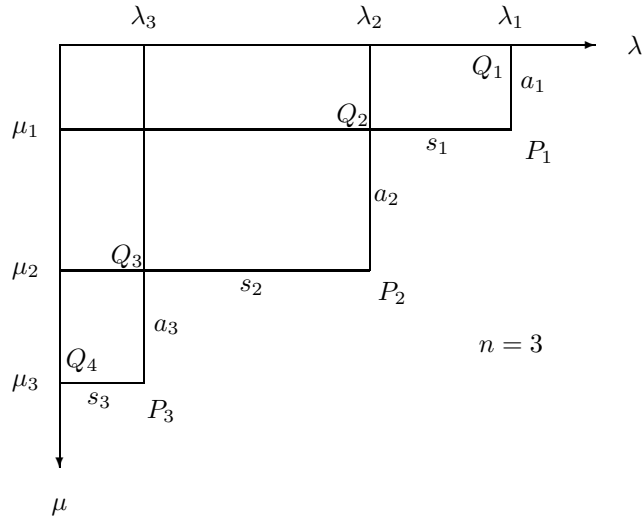
\unitlength=0.75mm
\linethickness{0.4pt}
\begin{picture}(122.00,126.00)(-15.00,005.00)
 \put(20.00,120.00){\line(1,0){85.00}}
\put(20.00,120.00){\line(0,-1){60}}
\put(100.00,120.00){\line(0,-1){15}}
\put(100.00,105.00){\line(-1,0){80}}
\put(75.00,120.00){\line(0,-1){40}}
\put(75.00,80.00){\line(-1,0){55}}
\put(35.00,120.00){\line(0,-1){60}}
\put(35.00,60.00){\line(-1,0){15}}
\put(38.00,55.00){\makebox(0,0)[cc]{$P_3$}}
\put(79.00,76.00){\makebox(0,0)[cc]{$P_2$}}
\put(105.00,101.00){\makebox(0,0)[cc]{$P_1$}}
\put(96.00,116.00){\makebox(0,0)[cc]{$Q_1$}}
\put(72.00,107.50){\makebox(0,0)[cc]{$Q_2$}}
\put(32.00,82.50){\makebox(0,0)[cc]{$Q_3$}}
\put(24.00,64.00){\makebox(0,0)[cc]{$Q_4$}}
\put(100.00,125.00){\makebox(0,0)[cc]{$\lambda _1$}}
\put(75.00,125.00){\makebox(0,0)[cc]{$\lambda _2$}}
\put(35.00,125.00){\makebox(0,0)[cc]{$\lambda _3$}}
\put(14.00,105.00){\makebox(0,0)[cc]{$\mu _1$}}
\put(14.00,80.00){\makebox(0,0)[cc]{$\mu _2$}}
\put(14.00,60.00){\makebox(0,0)[cc]{$\mu _3$}}
\put(87.00,102.00){\makebox(0,0)[cc]{$s_1$}}
\put(54.00,77.00){\makebox(0,0)[cc]{$s_2$}}
\put(27.00,57.00){\makebox(0,0)[cc]{$s_3$}}
\put(39.00,70.00){\makebox(0,0)[cc]{$a_3$}}
\put(78.00,93.00){\makebox(0,0)[cc]{$a_2$}}
\put(104.00,113.00){\makebox(0,0)[cc]{$a_1$}}
\put(100.00,120.00){\vector(1,0){15.00}}
\put(20.00,60.00){\vector(0,-1){15.00}}
\put(122.00,120.00){\makebox(0,0)[cc]{$\lambda$}}
\put(20.00,38.00){\makebox(0,0)[cc]{$\mu$}}
\put(100.00,67.00){\makebox(0,0)[cc]{$n=3$}}
\end{picture}

\vspace{-15mm}
 
In refs.\,\cite{KR,BR2} (see also \cite{KOS}) the transfer matrix for
a general Young diagram has been expressed through transfer matrices
for Young diagrams which consist of either a single row or a single
column. It is given by the determinants\footnote{We thank A.Kuniba for
  communication on this point.}: \beq T_{\lambda}^{\mu}(u)=\det_{1\leq
  i,j \leq \mu _n} \big ( T_{y_{i}-i+j}(u-\mu_n + \lambda_{1}-
y_{i}+i+j-1)\big )\,.
\label{det1}
\eeq
\beq
T_{\lambda}^{\mu}(u)=\det _{1\leq i,j \leq \lambda_1}
\big ( T^{y^\prime_{i}-i+j}(u-\mu_n + \lambda _{1}+
y^\prime_{i}-i-j+1)\big )\,.
\label{det2}
\eeq
Here $y_j$, is the length of the $j$-th row, $j=1,\ldots ,\mu_n $. 
Similarly, $y'_{k}$ are lengths of rows of the transposed diagram
 $Y '$ i.e. the diagram $Y$ reflected with respect to its
 main diagonal. The entries $T_{m}(u)$ ($T^{m}(u)$) of the
determinants are $T$-matrices corresponding to
the one-row (one column) Young diagram of length $m$.

We show that this function satisfies the 
bilinear equation:
\beq
T_{\lambda}^{\mu}(u-1)T_{\lambda}^{\mu}(u+1)-
T_{\lambda +1}^{\mu}(u)T_{\lambda -1}^{\mu}(u)=
\sum _{i=1} ^n T^{\mu +\theta ^i}_{\lambda +\theta ^{i+1}}(u)
T_{\lambda -\theta ^{i+1}}^{\mu -\theta^i}(u),
\label {HH}
\eeq
where
$$\lambda +1 \equiv (\lambda _1 +1,\ldots , \lambda _n +1)$$
and the step function
$\theta ^{i}=(\theta ^{i}_1,...,\theta ^{i}_n )$ is defined
by $\theta _{j}^{i}= 0$ if $j<i$ and
$\theta _{j}^{i}= 1$ if $j \geq i$, so
$$\mu + \theta ^i \equiv
(\mu _1,\ldots ,\mu _{i-1}, \mu _i +1,\mu _{i+1}
+1, \ldots ,\mu _n +1 ),$$
$$\lambda +\theta ^{i+1} \equiv (\lambda _1, \ldots ,\lambda _{i},
\lambda _{i+1} +1, \lambda _{i+2} +1, \ldots ,
\lambda _n +1 ).$$

In terms of the ladder coordinates 
($T_{s_1,...,s_n}^{a_1,...,a_n}(u)\equiv T_{\lambda }^{\mu}(u)$) 
the equation acquires the form
\begin{eqnarray}
&&T_{s_1,...,s_n}^{a_1,...,a_n}(u-1)T_{s_1,...,s_n}^{a_1,...,a_n}(u+1)-
T_{s_1,...,s_{n-1},s_n +1}^{a_1,...,a_n}(u)
T_{s_1,...,s_{n-1},s_n -1}^{a_1,...,a_n}(u)
\nonumber\\
&=&T^{a_1,...,a_n +1}_{s_1,...,s_n}(u)T^{a_1,...,a_n
-1}_{s_1,...,s_n}(u)
\nonumber \\
&+&\sum _{i=1}^{n-1}
T^{a_1,...,a_i +1,...,a_n}_{s_1,...,s_i -1,...,s_n +1}(u)
T^{a_1,...,a_i -1,...,a_n}_{s_1,...,s_i +1,...,s_n -1}(u).
\label{HHa}
\end{eqnarray}

In these relations the following boundary conditions are imposed:
\beq\label{bc} T_{s_1,...,s_n}^{a_1,...,a_n}(u)=0 \,\,\,\,\,\,
\mbox{if at least one of} \,\,\, a_i=-1.  \eeq Under this condition
eq.\,(\ref{HHa}) consists of diagrams with the number of corners $
\leq n$.  As soon as $a_i=0$, the term $T^{a_1,...,a_i
  -1,...,a_n}_{s_1,...,s_i +1,...,s_n -1}(u)$ is zero and drops out
from the sum.  All the remaining terms have $a_i=0$ and correspond to
diagrams with $n-1$ corners.

Eq.\,(\ref{HHa}) can be transformed to a higher Hirota form (\ref{H2})
by a linear change of variables. First of all, let us notice that the
equation actually depends on $n+2$ variables rather than $2n+1$
variables $a_i,s_i$. Indeed, the combinations
$$
q_i =\frac{1}{2}(a_i + s_i ), \;\;\;\;\;\;\;i=1, \ldots , n-1,
$$
are the same in all $T$-functions involved in eq. (\ref{HHa}) and,
therefore, can be considered as parameters.

It is convinient  to change the variables in two steps. First, introduce
the variables $p_j$, $j=0, 1 , \ldots , n+1$:
\begin{eqnarray}
&&p_0 = u,
\nonumber \\
&&p_i =\frac{1}{2} (a_i -s_i ), \;\;\;\;\;\;\;1\leq i \leq n-1,
\nonumber \\
&&p_n =a_n,
\nonumber \\
&& p_{n+1}=s_n -\frac{1}{2}\sum _{l=1}^{n-1}(a_l-s_l).
\label{change1}
\end{eqnarray}
Note that
$
\sum _{i=0}^{n+1} p_i = u+a_n + s_n \,.
$
In the new variables ($T(p_0 , p_1 , \ldots , p_{n+1})
\equiv T_{s_1,...,s_n}^{a_1,...,a_n}(u)$) the equation acquires the form
\beq
T(p_0 +1)T(p_0 -1) =\sum _{i=1}^{n+1}T( p_i +1) T(p_i -1),
\label{HHb}
\eeq
where we have dropped the variables that do not undergo shifts.

Next introduce the variables:
\beq
x_i =\frac{1}{2} \left ( p_i -\sum _{j=0, \neq i}^{n+1}p_j \right ),
\;\;\;\;\;\;\; i=0, 1 , \ldots , n+1.\label{change2}
\eeq
 In terms of the initial variables they are:
\begin{eqnarray}
&&x_0 =\frac{1}{2}(u-a_n - s_n ),
\nonumber \\
&&x_i =\frac{1}{2}(-u-a_n - s_n +a_i - s_i ),
\;\;\;\;\;\;\;1\leq i \leq n-1,
\nonumber \\
&&x_n =\frac{1}{2}(-u+a_n - s_n ),
\nonumber \\
&&x_{n+1} =\frac{1}{2}(-u+\lambda _{1} - \mu _{n}).
\label{change3}
\end{eqnarray}
Finally, passing to the function $\tau (x_0 ,  \ldots , x_{n+1})=
T(p_0 , \ldots , p_{n+1})$, we arrive at
eq.\,(\ref{H2}) with
$r=n+2$, $\alpha _{1}=-1$, $\alpha _{i}=1$, $i=2, 3 , \ldots , n+2$.

This proves that the bilinear fusion rules are equivalent
 to the higher Hirota equations. 

Let us notice that the bilinear relations for the 
$T$-matrices (\ref{HH}) are not unique. There exist other bilinear relations 
of a slightly different structure
\cite{Z}.

\section{General Solution to the Higher Hirota Equation (3.5)}

The higher Hirota equation with  particular boundary conditions (\ref{bc})
is solved explicitly along the lines of ref.\,\cite{KLWZ}.

For the series $A_{k-1}$, the number of rows in Young diagrams does
not exceed $k$. The boundary conditions (\ref{bc}) in this case are:
\beq T_{\lambda }^{\mu}(u)=0 \;\;\;\;\mbox{as}\;\;\; \mu _{n}>k
\;\;\;\mbox{or}\;\;\; \mu _{n}<0.
\label{bc1}
\eeq
In what follows, it is convenient 
 to set $\mu _{n+1}=k$.

The general solution depends on
 $(n+1)k$ arbitrary
functions $h_{l}^{j}(u)$, $l=1, \ldots , n+1$,
$j=1, \ldots , k$. Let us form the $k\times k$ matrix:
\beq
H_{i,j}^{(\mu , \lambda )}=h_{l(i)}^{j}(u-\lambda _{1}+\mu _{n}
+2\lambda _{l(i)} -2i +2),
\label{matH}
\eeq
where the function $l(i)$ is determined by the condition
$$
\mu _{l(i)-1}+1 \leq i \leq \mu _{l(i)}+1 \;\; ,
$$
and $i=1,2, \ldots , k$. This matrix has a horizontal strip structure:
the  $l$-th strip consists of
$\mu _{l}-\mu _{l-1}$ rows.

Then the general solution has the form:
\beq
T_{\lambda}^{\mu}(u)=\det _{1\leq i,j \leq k}
H_{i,j}^{(\mu , \lambda )}.
\label{gensol}
\eeq 
This form of the solution is convenient for deriving the Bethe
Ansatz equations \cite{BR2} for a general $A_{k-1}$ quantum integrable
model.  The derivation is along the lines of ref.\,\cite{KLWZ}.  Other
determinant representations also exist.

\section{Grassmannians and Pl\"ucker Relations}

The proof of the bilinear relations for the $T$-matrices is based on
the Pl\"ucker identities. They appear as relations between coordinates
on Grassmann manifolds (see
\cite{HT},\,\cite{Galois},\,\cite{GrifHar}).  The connection between
Hirota's bilinear equations and the geometry of Grassmann manifolds is
well known \cite{Sato},\,\cite{JimboMiwa},\,\cite{SW}.  The Grassmann
manifolds related to general solutions of a soliton equation are
infinite dimension.  Remarkably, the solutions of interest, specified
by the boundary conditions ({\ref{bc1}), correspond to {\it finite
    dimensional} grassmannians.  This allows one to write down a
  general solution in terms of determinants.  Numerous determinant
  formulas like (\ref{det1}), (\ref{det2}), (\ref{gensol}) may be
  obtained in this way.

The grassmannian ${\bf G}_{N+1}^{M+1}$ is a collection of
all $(M+1)$-dimensional linear subspaces of the complex
$(N+1)$-dimensional vector space ${\bf C}^{N+1}$. In particular,
${\bf G}_{N+1}^{1}$ is the complex projective space ${\bf P}^{N}$.
Let $X\in {\bf G}_{N+1}^{M+1}$ be such a $(M+1)$-dimensional
subspace spanned by vectors ${\bf x}^{(j)}=\sum _{i=0}^{N}x_{i}^{(j)}
{\bf e}^i$, $j=1,\ldots , M+1$, where the ${\bf e}^i$ are the basis vectors of
${\bf C}^{N+1}$. The collection of their coordinates form a
rectangular $(N+1)\times (M+1)$-matrix $x^{(j)}_i$.
Let us consider its $(M+1)\times (M+1)$ minors:
\beq\label{d}
\det _{pq}(x^{(q)}_{i_p})\equiv (i_{0},i_{1},\ldots ,i_{M}),\;\;\;\;\;
p,q=0,1, \ldots ,M\,,
\eeq
obtained by choosing $M+1$ lines $i_{0}, i_{1}, \ldots , i_{M}$.
These $C_{N+1}^{M+1}$ minors are called the {\it Pl\"ucker coordinates}
of $X$. They are defined up to a common scalar factor and provide the
{\it Pl\"ucker embedding} of the grassmannian ${\bf G}_{N+1}^{M+1}$
into the projective space ${\bf P}^d$, where $d=C_{N+1}^{M+1}-1$
($C_{N+1}^{M+1}$ is the bimomial coefficient).

The image of ${\bf G}_{N+1}^{M+1}$ in ${\bf P}^d$ is realized as
an intersection of quadratic curves. This means that the coordinates
$(i_{0},i_{1},\ldots ,i_{M})$ are not
independent but obey
the {\it Pl\"ucker relations} \cite{Galois},\,\cite{GrifHar}:
\beq
(i_{0},i_{1},...,i_{M})(j_{0},j_{1},...,j_{M})= \sum_{p=0}^{M}
(j_{p},i_{1},...,i_{M})(j_{0},...j_{p-1},i_{0},j_{p+1}...,j_{M})
\label{p-relation}
\eeq
for all $i_p , j_p$, $p=0,1, \ldots , M$.

\section{The Pl\"ucker Relations and Fusion Rules}

Here we outline the proof of eq.\,(\ref{HH}). It turns out that the
bilinear fusion rules are realized as the Pl\"ucker relations. In
order to compare them let us put $i_p = j_p$ for $p\ne 0,1, \ldots ,
n$ in (\ref{p-relation}). Then all terms but the first $n+1$ terms in
the r.h.s.  of (\ref{p-relation}) vanish. Then the Pl\"ucker relations
read \beq \phantom{a}[i_{0},i_{1},\ldots ,i_{n}] \cdot
[j_{0},j_{1},j_{2}, \ldots ,j_{n}]= \sum
_{p=0}^{n}[j_{p},i_{1},i_{2},\ldots ,i_{n}]\cdot [j_{0}, \ldots ,
j_{p-1}, i_{0},j_{p+1},\ldots ,j_{n}],
\label{useful}
\eeq
where we denote
$$ \phantom{a}[i_{0}, \ldots , i_{n}]\equiv
(i_{0}, \ldots , i_{n-1}, i_{n}, \ldots , i_{M}),
$$
in order to stress that all arguments with a 
subscript greater than $n$ are the same in all terms of 
eq.\,(\ref{useful}). Only the first $n$ arguments are shown explicitly.

Given a Young diagram $Y(\lambda , \mu )$ and the corresponding
function $T_{\lambda }^{\mu }(u)$, introduce a rectangular
$(\mu _{n} +n +2 )\times (\mu _{n}+1)$-matrix $M_{i,j}$,
$i=-1 , 0, 1 , \ldots , \mu _{n}+n$,
$j=1 , 2, \ldots , \mu _{n}+1$. Its matrix elements are:
\beq\label{x}
M_{-1, j}=\delta _{j, \mu _{n}+1 },\;\;\;\;\;\;\;
M_{0, j}=\delta _{j, 1 }
\eeq
\beq
M_{i,j}=T_{\lambda _{l(i)}-i+j +l(i) -1}(u+\lambda _{1}-\mu _{n}
-\lambda _{l(i)} +i+j -l)
\label{matM}
\eeq
where $l(i)$ is given by
$$\mu _{l(i)-1}+l(i) \leq i \leq \mu _{l(i)}+l(i),\;\;\;\;\;\;\;\;
l=1,2, \ldots , n
$$
and $\delta _{i,j}$ is Kronecker's symbol.
The matrix consists of $n+1$ horizontal strips. Each strip except the
first one consists of 
$\mu_l-\mu_{l-1}$ rows.
 The first strip has only two rows (\ref{x}). 
 
 Let us apply the Pl\"ucker relation (\ref{useful}) to the minors of
 this matrix. Choose $i_{p}$ ($j_{p}$) to be the first (the last) row
 of the $p$-th strip:
$$i_{0}=-1,\;\; j_{0}=0,\;\;
i_{l}=\mu _{l-1}+l , \;\;
j_{l}=\mu _{l}+l , \;\;\;\;\;\;\;l=1, \ldots , n.
$$
Now, using eq.\,(\ref{det1}), we identify the minors with 
the $T$-matrices:
\begin{eqnarray}
&&[i_0, i_1,\ldots ,i_n]=\omega (-1)^{\mu _{n}}
T_{\lambda }^{\mu}(u),
\nonumber \\
&&[j_0, j_1,\ldots ,j_n]=\omega '
T_{\lambda }^{\mu}(u+2),
\label {pluk1}
\end{eqnarray}
where $\omega , \omega '$ are irrelevant sign factors such that
$\omega \omega ' =(-1)^{\mu _{n}+n}$.
Computing the remaining minors, we obtain :
\begin{eqnarray}
&&[j_0, i_1, \ldots , i_n]=\omega T_{\lambda +1}^{\mu}(u+1),
\nonumber \\
&&[i_0, j_1,\ldots ,j_n]=
\omega ' (-1)^{\mu _n} T_{\lambda -1}^{\mu}(u+1),
\nonumber \\
&&[j_p, i_1,\ldots ,i_n]=\omega
(-1)^{\mu _p} T^{\mu +\theta ^p}_{\lambda +\theta ^{p+1}}(u+1),
\nonumber \\
&&[j_0,\ldots , j_{p-1},i_0,j_{p+1},\ldots ,j_n]=
\omega ' (-1)^{\mu _p +\mu _n}
T^{\mu -\theta ^p}_{\lambda -\theta ^{p+1}}(u+1).
\label{pluk3}
\end{eqnarray}
Since the minors of the matrix (\ref{matM}) obey the Pl\"ucker
relations (\ref{useful}), the $T$-matrices satisfy eq.\,(\ref{HH}).

The proof of  eq.\,(\ref{gensol}) is 
similar.

\section{Conclusion}

We have shown that the fusion rules of quantum integrable models have
the form of an integrable dynamical system on a finite dimensional
Grassmann manifold.  Namely, the fusion relations are identical to the
hierarchy of Hirota difference equations with open boundary
conditions.

This equation, supplemented by analyticity requirements \cite{KLWZ}
completely determines the spectrum of the quantum system, i.e. reveals
the Bethe Ansatz equations.

In this note we just observed a connection between the fusion
procedure and Grassmannian geometry. It seems to be an intimate one
and is also expected for the fusion of conformal field theories. A
deeper understanding of this connection would be desirable.

\section{Acknowledgements}

We thank H.Awata, S.Khoroshkin and A.Kuniba for discussions
 and J.Talstra for help. 
 The work of O.L. was supported by the  MRSEC NSF grant DMR 9400379.
The work of A.Z. was supported in part by RFBR grant
95-01-01106, by ISTC grant 015 and also by NSF grant DMR-9509533.
 P.W. was supported by NSF grant DMR-9509533. 
P.W and A.Z thank the Institute for Theoretical Physics in Santa Barbara for
its hospitality in April 1997 when this paper was completed.


\begin{thebibliography}{99}

\bibitem{FT}
L.D.Faddeev and L.A.Takhtadjan,
{\it Quantum inverse scattering method
and the $XYZ$ Heisenberg model},
Uspekhi Mat. Nauk {\bf 34:5} (1979) 13-63.

\bibitem{fusion} P.P.Kulish, N.Yu.Reshetikhin and E.K.Sklyanin,
{\it Yang-Baxter equation and representation theory: I},
Lett. Math. Phys. {\bf 5} (1981) 393-403;
P.P.Kulish and E.K.Sklyanin, {\it Quantum spectral
transform method. Recent developments}, Lecture Notes in Physics {\bf 151}
61-119, Springer, 1982.

\bibitem{KP} A.Kl\"umper and P.Pearce, {\it Conformal weights of
RSOS lattice models and their fusion hierarchies}, Physica
{\bf A183} (1992) 304-350.

\bibitem{Kuniba} A.Kuniba, T.Nakanishi and J.Suzuki,
{\it Functional relations in solvable lattice models, I: Functional
relations and representation theory, II: Applications}, Int. Journ. Mod.
Phys. {\bf A9} (1994) 5215-5312.

\bibitem{KLWZ} I.Krichever, O.Lipan, P.Wiegmann and A.Zabrodin,
{\it Quantum integrable models and discrete classical Hirota
equations}, preprint ESI 330 (1996), hep-th/9604080, to appear
in Commun. Math. Phys.

\bibitem{Hirota1} R.Hirota, {\it Discrete analogue of a generalized
Toda equation}, Journ. Phys. Soc. Japan {\bf 50} (1981) 3785-3791.

\bibitem{Sato} M.Sato, {\it Soliton equations as dynamical systems
on infinite dimensional Grassmann manifolds}, RIMS Kokyuroku
{\bf 439} (1981) 30-46.

\bibitem{Miwa2}  E.Date, M.Jimbo and T.Miwa, {\it Method for
generating discrete soliton equations I, II}, Journ. Phys. Soc. Japan
(1982) 4116-4131.

\bibitem{OHTI} Y.Ohta, R.Hirota, S.Tsujimoto and T.Imai,
{\it Casorati and discrete Gram type determinant representations
of solutions to the discrete KP hierarchy}, Journ. Phys. Soc. Japan
{\bf 62} (1993) 1872-1886.

\bibitem{KR} P.P.Kulish and N.Yu.Reshetikhin,
{\it On $GL_{3}$-invariant solutions of the Yang-Baxter equation
and associated quantum systems}, Zap. Nauchn. Sem. LOMI {\bf 120}
(1982) 92-121 (in Russian), Engl. transl.: J. Soviet Math. {\bf 34}
(1986) 1948-1971;
N.Yu.Reshetikhin, {\it The functional equation
method in the theory of exactly soluble quantum systems},
Sov. Phys. JETP {\bf 57} (1983) 691-696.

\bibitem{BR2} V.Bazhanov and N.Reshetikhin, {\it Restricted solid on solid
models connected with simply laced algebras and conformal field theory},
Journ. Phys. {\bf A23} (1990) 1477-1492.

\bibitem{KOS} A.Kuniba, Y.Ohta and J.Suzuki, {\it Quantum
Jacobi-Trudi and Giambelli formulae for $U_q (B_{r}^{(1)}$ from
analytic Bethe ansatz}, preprint, hep-th/9506167.

\bibitem{Z} A.Zabrodin, {\it Discrete Hirota's equation in
quantum integrable models}, preprint ITEP-TH-44/96 (1996),
hep-th/9610039.

\bibitem{HT} W.V.D.Hodge and D.Pedoe, {\it Methods of algebraic geometry},
volume I, Cambridge University Press, Cambridge, 1947.

\bibitem{Galois} J.W.P.Hirschfeld and J.A.Thas, {\it General Galois
geometries}, Claredon Press, Oxford, 1991.

\bibitem{GrifHar} P.Griffiths and J.Harris, {\it Principles of
algebraic geometry}, A Wiley-Interscience Publication,
John Wiley {\&} Sons, 1978.

\bibitem{JimboMiwa} M.Jimbo and T.Miwa, {\it Solitons and infinite
dimensional Lie algebras}, Publ. RIMS, Kyoto Univ. {\bf 19} (1983)
943-1001.

\bibitem{SW} G.Segal and G.Wilson, {\it Loop groups and equations
of KdV type}, Publ. IHES {\bf 61} (1985) 5-65.



\end{thebibliography}
\end{document}